\pdfoutput=1
\documentclass[11pt]{article}

\usepackage[final]{acl}

\usepackage{times}
\usepackage{latexsym}
\usepackage{textcomp}
\usepackage[T1]{fontenc}
\usepackage[utf8]{inputenc}
\usepackage{microtype}
\usepackage{inconsolata}
\usepackage{booktabs}
\usepackage{xspace}
\usepackage{amsmath}
\usepackage{amssymb}
\usepackage{mathtools}
\usepackage{tcolorbox}
\usepackage{xspace}
\usepackage{caption}
\usepackage{subcaption}
\usepackage{balance}
\usepackage[hang,flushmargin]{footmisc}
\usepackage{multirow}

\newcommand{\ourmodel}{DSE}

\title{Unifying Multimodal Retrieval via Document Screenshot Embedding}

\author{
Xueguang Ma\quad
Sheng-Chieh Lin\quad
Minghan Li \quad
Wenhu Chen\quad
Jimmy Lin\\
[1ex]
David R.\ Cheriton School of Computer Science, University of Waterloo \\
[1ex] \texttt{\{x93ma, s269lin, m692li, wenhuchen, jimmylin\}@uwaterloo.ca}
}
\begin{document}
\maketitle
\begin{abstract}
In the real world, documents are organized in different formats and varied modalities.
Traditional retrieval pipelines require tailored document parsing techniques and content extraction modules to prepare input for indexing.
This process is tedious, prone to errors, and has information loss.
To this end, we propose \textit{\underline{D}ocument \underline{S}creenshot \underline{E}mbedding} (\ourmodel), a novel retrieval paradigm that regards document screenshots as a unified input format, which does not require any content extraction preprocess and preserves all the information in a document (e.g., text, image and layout).
\ourmodel{} leverages a large vision-language model to \textit{directly} encode document screenshots into dense representations for retrieval. 
To evaluate our method, we first craft the dataset of Wiki-SS, a 1.3M Wikipedia web page screenshots as the corpus to answer the questions from the Natural Questions dataset.
In such a text-intensive document retrieval setting, \ourmodel{} shows competitive effectiveness compared to other text retrieval methods relying on parsing. 
For example, \ourmodel{} outperforms BM25 by 17 points in top-1 retrieval accuracy.
Additionally, in a mixed-modality task of slide retrieval, \ourmodel{} significantly outperforms OCR text retrieval methods by over 15 points in nDCG@10.
These experiments show that \ourmodel{} is an effective document retrieval paradigm for diverse types of documents.
Model checkpoints, code, and Wiki-SS collection are released at \url{http://tevatron.ai}.

\end{abstract}

\section{Introduction}

\begin{figure*}
    \centering
    \includegraphics[width=\textwidth]{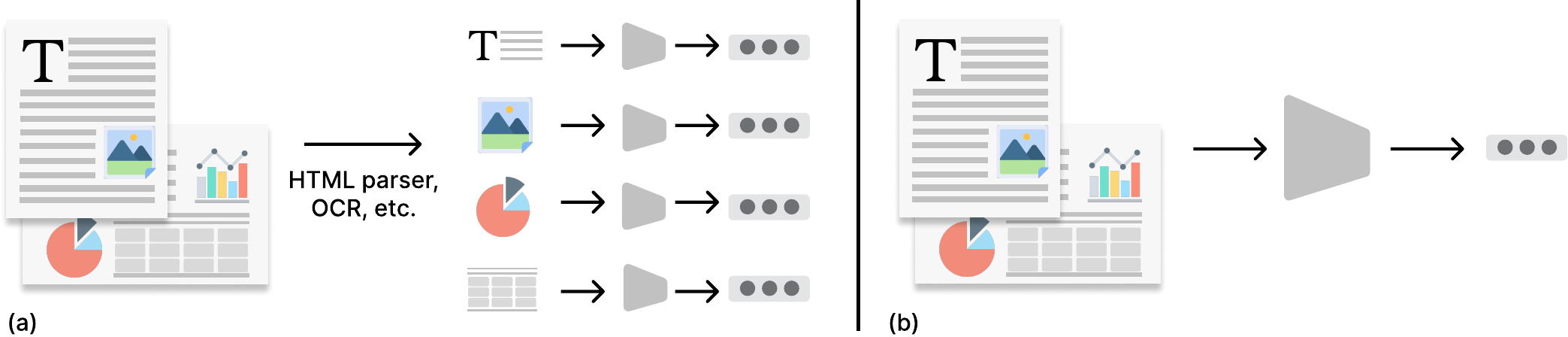}
    \caption{Comparison between (a) existing document retrieval paradigm and (b) our proposed paradigm.
    \ourmodel{} bypasses the document parsing and content extraction process, directly encoding the original appearance of documents with multimodal contents into a dense representation for indexing}
    \label{fig:paradigm}
\end{figure*}

Information retrieval systems help users access external information from documents in varied modalities, including text, images, charts, and tables.
As shown in Figure~\ref{fig:paradigm}(a), existing document retrieval paradigms typically process these modalities separately.
For example, traditional lexical retriever BM25~\cite{Robertson2009ThePR} or neural retrievers such as DPR~\cite{karpukhin-etal-2020-dense} rely on extracted text contents from documents.
Recent multimodal retrieval~\cite{yang2023atomic, wei2023uniir} leverage both processed text and image units to broaden the scope of retrieval, thus supporting text-image tasks.
% , aiming to unify multimodal representation learning.

However, the existing retrieval paradigms lack a unified encoding process across modalities, leading to two underlying issues. 
Firstly, preprocessing is not a trivial effort. 
Specialized processing is required to handle various document types and content modalities, and they are often imperfect. 
For instance, HTML files in the wild can present significant complexity due to their varied structures, making it difficult for a single tool to parse all information accurately. 
Similarly, slides and PDFs often require OCR models to extract text and handle other content types like tables and figures separately~\cite{layoutlmv3,SlideVQA2023}.
Managing these diverse modalities separately is tedious, and precisely dealing with the long-tailed document appearances in the real world is often impractical.
Secondly, this process ``breaks'' the original appearance of the document, disrupting its visual context and layout integrity. The visual presentation of a document can convey essential information that is difficult to capture through content extraction alone.
For example, in addition to the contents of texts and images, the size and position of these elements in a document may encode the importance of the information they contain~\citep{layoutlm, layoutlmv3}.

To tackle the aforementioned issues, we 
introduce \textit{\underline{D}ocument \underline{S}creenshot \underline{E}mbedding} (\ourmodel), a new information retrieval paradigm that unifies the varied formats and modalities in a single form for direct document encoding and indexing: screenshots. 
In contrast to using various tools to extract texts and images from documents in different formats, screenshots are easy to obtain and all the information in the documents are visually preserved. 

As illustrated in Figure~\ref{fig:paradigm}(b), 
\ourmodel{} directly encodes the screenshot of any given document into a dense representation through a large vision-language model.
During search, a user's query is encoded by a language model to locate the nearest document embeddings.
We conduct empirical studies to demonstrate that \ourmodel{} is effective for document retrieval.
Specifically, we conduct experiments on two types of document retrieval settings: text-intensive and text-image-mixed.
For the text-intensive case, we collect 1.3 million Wikipedia web page screenshots as our corpus and fine-tune a large vision-language model as a bi-encoder to conduct dense retrieval on questions in the NQ dataset~\cite{kwiatkowski-etal-2019-natural}. 
Experimental results show that \ourmodel{} outperforms the traditional text-based retrieval method BM25 by 17 points in top-1 retrieval accuracy on NQ questions and is competitive with text-based dense retrieval methods in a text-oriented evaluation.
This experiment indicates that \ourmodel{} can sufficiently encode the textual information in a screenshot.
As for the image-text mixed setting, we use slide retrieval.
We turn the existing SlideVQA~\cite{SlideVQA2023} dataset into an open-domain retrieval setting, where models are required to retrieve relevant slides from a pool of 50k slides for given questions. 
Results show that \ourmodel{} outperforms all text-based retrieval methods which rely on OCR (including BM25 and dense text retrieval) by over 15 points in nDCG@10.

\section{Related Work}
\subsection{Neural Document Retrieval}
Traditional document retrieval methods such as TF-IDF and BM25~\cite{Robertson2009ThePR} represent text as bag-of-words representations and conduct efficient search over an inverted index.
Recent neural retrieval methods represented by DPR~\cite{karpukhin-etal-2020-dense}, proposed to finetune pretrained neural networks such as BERT~\cite{devlin-etal-2019-bert} to encode query and document separately into dense semantic vectors in a bi-encoder architecture.
The effectiveness of text dense retriever has been boosted in recent years by various training strategies such as data augmentation~\cite{xiong2021approximate, lin-etal-2023-train, bge_embedding}, pretraining~\cite{izacard2021contriever, gao-callan-2022-unsupervised, wang-etal-2023-simlm}, distillation~\cite{lin-etal-2021-batch, ren-etal-2021-rocketqav2} and instruction tuning~\cite{su-etal-2023-one, asai-etal-2023-task}.
With the growth of large language models, finetuning an LLM-based text encoder demonstrated further improvement in both in-domain and out-domain retrieval effectiveness~\cite{rankllama, wang2023improving, muennighoff2024generative, lee2024nvembed}.

Besides text retrieval, prior multi-modal retrieval studies~\cite{wei2023uniir, koukounas2024jina} have explored retrieval across various combinations of text and image inputs for queries and documents.
These approaches aim to bridge the gap between different modalities, enabling more comprehensive retrieval systems.
Existing text and multi-modal retrieval works assume that the datasets are well pre-processed, where text and image data are carefully extracted and organized for model inputs.
However, this is not always true in real-world scenarios where documents are often unstructured and diverse.
In this work, we consider the document retrieval tasks that begin with the original look of documents.

\subsection{Large Vision-Language Model}
Large language models (LLMs) like GPT-4~\cite{openai2024gpt4} and LLaMA~\cite{touvron2023llama}, pre-trained on massive corpora and fine-tuned to follow user instructions, have shown success in various natural language generation tasks~\cite{wei2022chain}.
Recent advancements have integrated vision capabilities into LLMs, enabling them to process both text and images simultaneously. Commercial models like GPT-4V~\cite{openai2024gpt4} and open-source models such as LLaVA~\cite{liu2023visual} exhibit strong performance. 
Building upon LLaVA, recent works such as LLaVA-NEXT~\cite{liu2024llavanext}, Idefics2~\cite{laurençon2024matters}, and Phi-3-vision~\cite{abdin2024phi3} have further improved performance.
They enable the processing of higher-resolution images and handle more challenging vision-language tasks, such as OCR~\citep{liu2024llavanext, liu2024hidden}.
Inspired by the capabilities of large vision-language models, our work pioneers its application in document retrieval tasks.

\begin{figure*}
    \centering
    \includegraphics[width=\textwidth]{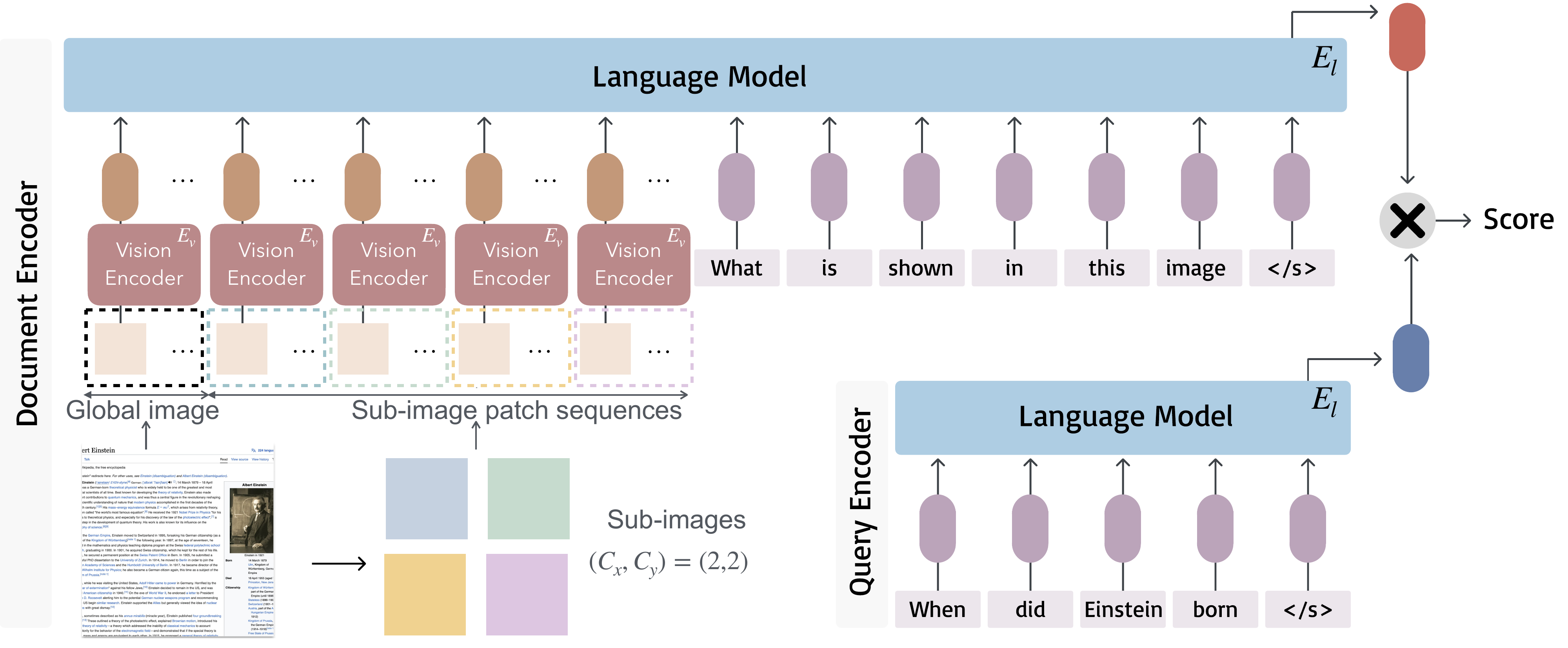}
    \caption{Overview of \ourmodel{} encoder architecture. \ourmodel{} adopts a bi-encoder architecture, where the document tower encodes the document screenshot into dense vector by taking vision input and the query tower encodes the query by taking text input. 
    Document and query encoders share the same language model.
    }
    \label{fig:enter-label}
\end{figure*}

\subsection{Document Retrieval Datasets}
Commonly used text retrieval datasets such as MS MARCO~\cite{bajaj2018ms}, Wikipedia-NQ~\cite{karpukhin-etal-2020-dense}, and BEIR~\cite{thakur2021beir} are released in well-preprocessed text contents. Similarly, multi-modal retrieval datasets like AToMIC~\cite{yang2023atomic} and m-BEIR~\cite{wei2023uniir} have text and images extracted from their sources and separately stored.

On the other hand, existing datasets designed for question-answering tasks based on document images include DocVQA~\cite{mathew2021docvqa}, VisualMRC~\cite{VisualMRC2021}, WebSRC~\cite{chen-etal-2021-websrc}, and InfographicVQA~\cite{Mathew_2022_WACV}. These datasets contain document images paired with questions, focusing on reading comprehension evaluation where a ground truth document image is provided for each question. Besides, the image pools in these datasets are relatively small, comprising only a few thousand images.

Therefore, to fairly evaluate multi-modal document retrieval in a large scale, we craft a text-intensive image corpus called Wiki-SS, containing 1.3 million Wikipedia page screenshots.
Additionally, we convert SlideVQA~\cite{SlideVQA2023} dataset, a visual QA dataset, into an open-domain slide retrieval dataset, consisting of 50K slides.

\section{Method}
\subsection{Task Definition}
Given a query $Q$ and a corpus $\mathcal{C}$ consisting of documents $\{D_1, D_2, ..., D_n\}$, the task of document retrieval is to identify the $k$ documents that are most relevant to the query $Q$, with $k\ll n$. This relevance is determined using a similarity metric $\textrm{Sim}(Q, D)\in \mathbb{R}$. 
Note that in this work, the screenshotted ``document'' is a complete information snippet (e.g. a web article, a PDF page).
This is different from some of the previous retrieval work, where the term ``document'' denotes arbitrary information snippets like sentences or passages. 
For queries, we only consider the text inputs similar to the traditional search setting. 
We leave the exploration of handling image queries for future work.

\subsection{Document Screenshot Embedding}
\label{sec:method}

We adopt a bi-encoder architecture for dense retrieval, where a document screenshot and user text query are encoded into dense vectors using a vision and text encoder, respectively. 
We can naively apply the vision and text encoders from CLIP~\cite{radford2021learning} to our task; however, in our experiment, we observe that the vision encoder cannot encode screenshots with more fine-grained information; thus, we propose to use large vision language models as the document screenshot encoder. 

\paragraph{Visual Encoder}
When a document screenshot $D$ is provided, it is first processed by a vision encoder $E_v$ to generate a sequence of latent representations. 
The length of the sequence is determined by the image tokenizer of the vision encoder. 
We take \texttt{
clip-vit-large-patch14-336}\footnote{\href{https://huggingface.co/openai/clip-vit-large-patch14-336}{ViT-Large}} as an example. 
Any given screenshot is first converted to an image with $336\times336$ pixels and then divided into $24\times24$ patches (i.e., 576 patches in total), each of which consists of $14\times14$ pixels. 
Each patch is flattened and mapped to a patch embedding with a trainable linear projection. 
The patch embeddings are encoded into latent representations with a vision encoder.
However, if a screenshot contains many texts (e.g., Wikipedia webpage), the 576 patch latent embeddings may not capture the fine-grained textual information in the screenshot.   

\paragraph{Vision Language Model}
To address the above issue, we leverage a large vision language model, Phi-3-vision,\footnote{\href{https://huggingface.co/microsoft/Phi-3-vision-128k-instruct}{Phi-3-vision}} which uses the same image tokenizer from \texttt{
clip-vit-large-patch14-336} but can represent an image with more patches by cropping it into sub-images.
For example, given a screenshot, we can choose to divide it into $(C_x\times24)\times(C_y\times24)$ patches. 
The given screenshot is converted to an image with $(C_x\times336)\times(C_y\times336)$ pixels and cropped into $C_x \times C_y$ sub-images, each of which has $336\times336$ pixels. 
Similarly, each sub-image is encoded into 576 patch latent representations independently. 
Note that Phi-3-vision further converts the whole screenshot into $336\times336$ pixels and encodes them into an additional 576 patch latent representations to capture the global information, resulting in $(C_x \times C_y + 1) \times 576$ patch latent representations in total, as depicted in left side of Figure~\ref{fig:enter-label}.  
Also, every four patch latent representations are concatenated and projected into one embedding for language model inputs. 
This process yields $(C_x \times C_y + 1) \times \frac{576}{4}$ patch latent embeddings as the input for the language model $E_l$. 
In Section ~\ref{sec:impacts_of_patch}, we will show that encoding a screenshot into more patch latent embeddings (increasing $C_x$ and $C_y$) helps capture more fine-grained information in the screenshot but sacrifices screenshot document encoding efficiency.

The encoded patch latent embeddings are concatenated with a text prompt as the input to the subsequent language model: \textit{``<s><img> What is shown in this image?</s>''}. 
Here, the <img> token is a special placeholder token and is replaced by the sequence of patch latent embeddings from the vision encoder. 
To aggregate sequence information using a language model with uni-directional attention, following previous work in text retriever~\cite{rankllama}, we use the embedding of the end-of-sequence token </s> from the last hidden state as the document screenshot embedding:
\[
V_d = E_l(E_v(D),\text{prompt})[-1]
\]

\paragraph{Contrastive Learning}
The similarity between the query and the document is computed as the cosine similarity between their embeddings:
\[
\text{Sim}(Q, D) = \frac{V_q^\top V_d}{\left \| V_q \right \|\cdot \left \| V_d \right \|}.
\]

During training, our embedding model is optimized using the InfoNCE loss:
\begin{equation}\label{eq:dpr_obj}
\centering
\begin{split}
    &\mathcal{L}(Q, D^+, D_{\text{N}}) \nonumber = -\log p(D=D^+\mid Q) \\
    & = -\log \frac{\exp(\text{Sim}(Q, D^+)/\tau)}{\sum\limits_{D_i \in \{D^+ \} \cup D_N} \exp(\text{Sim}(Q, D_i)/\tau)},
\end{split}
\end{equation}

\noindent where $D^+$ denotes the positive document. $D_\text{N}$ represents a set of negative documents that are irrelevant to the query $Q$, including hard negatives and in-batch negatives.
$\tau$ is a temperature parameter set to 0.02 in our experiments.
Note that we only consider text queries, which are directly input to the language model using template \textit{f``<s>\{\text{query}\}</s>''} and the last hidden state of \textit{</s>} is used as the query embedding, $V_q = E_l(Q)[-1]$.

\section{Experiment Setup}
\subsection{Web-Page Retrieval}\label{sec:exp:web}
\paragraph{Dataset} We construct the Wiki-SS dataset, using the Selenium Python toolkit\footnote{\url{https://pypi.org/project/selenium/}} to access English Wikipedia pages through URLs and automatically take screenshots.
The screenshots are taken with a window size of $980\times980$ pixels to ensure adequate coverage of the core content.
The screenshot creation process is conducted over a span of four days, from May 20 to May 23, 2024. 
Note that storing the entire collection of Wikipedia screenshots would require over 2TB of storage in PNG format.
In order to make Wiki-SS more manageable for research purposes, we downsize the corpus by filtering out the web pages which are considered ``easy negative samples'' for all the questions in the train, dev and test sets from Natural Questions~\cite{kwiatkowski-etal-2019-natural}.
Specifically, we perform BM25 search for each question to retrieve the top 50 documents over the text corpus.
The retrieved documents are pooled together as our final corpus.
Note that we concatenate each question and its corresponding ground truth answers as a query for BM25 search.
Although BM25 is a relatively weak retriever, including the target answer in the query for lexical search ensuring that positive and hard negative documents for each question are included in the downsized corpus.
As a result, we obtain a collection of 1,267,874 Wikipedia screenshots for our experiments.

To compare with text-based retrieval baselines, we create a text version Wikipedia collection which mirrors the collection of Wiki-SS.
Given the significant updates and changes to Wikipedia pages over time, the existing Wikipedia dumps~\cite{karpukhin-etal-2020-dense, atlas} cannot be used as a fair comparison. 
Thus, we re-process the Wikipedia text contents based on the May 20, 2024 dump\footnote{\url{https://huggingface.co/datasets/legacy-datasets/wikipedia}
} using Wikipedia parsing tool \texttt{mwparserfromhell}.
For each document in the text corpus, we use the first 500 words of each document, mirroring the corpus in Wiki-SS, where each screenshot covers only the first-page content.
For more details, please see Appendix~\ref{app:wiki-ss}.

\paragraph{Training Data}
We create the training data by taking the questions in the NQ train split as queries and using BM25 to retrieve the top-50 relevant documents over the text corpus for each question.
A document candidate (either in screenshot or text) is considered positive when the corresponding text contains the answers for the question.
Otherwise, the document is considered a hard negative candidate.
We drop the training example if either the positive or negative candidate list is empty, resulting in 49,095 training examples of triplets of query, positive documents and hard negative documents.

\paragraph{Evaluation}
We evaluate the in-domain effectiveness of retrievers using the 3,610 NQ test set questions. Consistent with previous practices in evaluating retrieval effectiveness on QA datasets~\cite{karpukhin-etal-2020-dense}, we use top-k retrieval accuracy as the metric. 
A question is considered correctly answered if one of the candidate documents contains an exact match of the answer string in the corresponding text content. 
The original NQ datasets contain both short answers and long answers for the questions~\cite{kwiatkowski-etal-2019-natural}, we follow the same method for computing exact match accuracy as \citet{karpukhin-etal-2020-dense}, where the short answers are the target.

\begin{table*}[t]
\centering
\begin{tabular}{l|c|cccc|cccc}
\toprule
\textbf{Retriever} & \textbf{Document}& \multicolumn{4}{c|}{\textbf{NQ}} & \multicolumn{2}{c}{\textbf{SlideVQA-open}} \\
\textbf{} & & Top 1 & Top 5 & Top 10 & Top 20 & nDCG@10 & Recall@10\\
\midrule
BM25 & \multirow{4}{*}{Text} & 29.5 & 52.6 & 61.3 & 67.3 & 55.8 & 63.7  \\
DPR &  & 42.3 & 63.9 & 69.7 & 74.3 & 47.4 & 57.9\\
E5 &  & 47.6 & 68.6 & 73.8 & 77.6 & 59.3 & 69.6 \\
Phi-3 &  & 50.6 & 70.9 & 75.8 & 79.5 & 59.0 & 69.5 \\
\midrule
CLIP & \multirow{2}{*}{Screenshot} & 35.1 & 57.7 & 64.8 & 71.2 & 61.7 & 74.7 \\
\ourmodel{} &  & 46.2 & 68.5 & 73.7 & 77.6 & 75.3 & 84.6 \\
\bottomrule
\end{tabular}
\caption{Supervised retrieval effectiveness comparison. \ourmodel{} and CLIP directly encode document screenshots while the other text-based retrieval models encode the extracted text from documents.}
\label{tab:main}
\vspace{-0.2cm}
\end{table*}

\subsection{Slide Retrieval}
\paragraph{Dataset} The original SlideVQA~\cite{SlideVQA2023} data is designed for document visual question answering. 
It contains 14.5k QA pairs and 52k slide images in total. The images contain various text formats, layouts, and visual content such as plots and charts. 
Given a question, the original task is to select the most relevant slides among the same deck with up to 20 slides and then answer the question based on the selected slides. 
The document selection process is in the form of reranking and classification.
In order to support the evaluation of document retrieval, we modify the SlideVQA to an open-domain retrieval task, where the task is to retrieve $k$ most relevant slide from the entire pool of slide images.
After our processing (e.g. removing the slides that fail to download, and questions that do not have evidence slides available), SlideVQA-open contains 50,714 slide images (screenshots) in its corpus.
We also create a corresponding text-based corpus for comparison with text retrievers using \texttt{pytesseract} OCR toolkit to extract text from every slide deck.

\paragraph{Training Data}
We create the training data based on the original train split of SlideVQA, the annotated evidence slides for a given question are considered positive documents, and the other slides within the same deck are considered as hard negative documents.
This process leads to 10,290 training examples in total.

\paragraph{Evaluation}
We construct the SlideVQA-open evaluation set using the 2,136 questions in the test set of SlideVQA. 
We evaluate the models' retrieval effectiveness using nDCG@10 and Recall@10.
In the following sections, mentions of SlideVQA refer to the open-domain retrieval setup.

\subsection{Implementation Details}
We implement \ourmodel{} by modifying the Tevatron toolkit~\cite{tevatron}, with the model initialized using Phi-3-vision~\cite{abdin2024phi3}, one of the state-of-the-art open-source large vision-language models with 4 billion parameters.
This model is recognized for its effective and efficient trade-off in performance.
To train the model, we employ memory-efficient techniques such as LoRA~\cite{hu2022lora}, FlashAttention~\cite{dao2023flashattention2}, and DeepSpeed~\cite{deepspeed}. 
The model is trained with a batch size of 128 for one epoch on Wikipedia webpage retrieval and trained with a batch size of 64 for two epochs for slide retrieval. 
The model weights are shared between the language models for document screenshot and query encoding.  
In both tasks, each training query is paired with one positive document and one hard negative document. 
We set $(C_x, C_y) = (4, 4)$ by default; that is, the document screenshots are resized to $1344\times1344$ pixels and cropped into $4\times4$ sub-images. The training process is conducted on two A100 80GB GPUs.
During inference, the embeddings are indexed using a Flat Faiss index~\cite{douze2024faiss} for exact nearest neighbor search.

\subsection{Baselines}
We compare \ourmodel{} against the following document retrieval methods based on text input: (1) BM25: a traditional text retriever based on lexical representation. 
(2) DPR: we follow the same setting as the DPR work~\cite{karpukhin-etal-2020-dense}, initializing dense retriever with BERT-base, and finetuning the model on our training data based on text input. 
(3) E5: similar to DPR, we finetune the unsupervised E5-base model~\cite{wang2022text}, which has BERT further pretrained with contrastive learning based on web data. 
(4) Phi-3: we use the same model initialization and configuration as \ourmodel{} but only fine-tune the component of the language model as a text-based dense retriever.
Additionally, we compare the fine-tuned CLIP model, whose image encoder is also initialized by ViT-large (the same as \ourmodel{}) but only supports a fixed length of patch sequence; i.e., $(C_x, C_y) = (1, 1)$. 
Please see Appendix~\ref{sec:hyper} for the detailed hyper-parameters of \ourmodel{} and baselines.

\section{Experimental Results}
\subsection{Supervised Retrieval Effectiveness}
Table~\ref{tab:main} presents the models' retrieval effectiveness in the supervised setting, where models are fine-tuned on NQ or SlideVQA training queries and evaluated on the corresponding evaluation set. 
For the Wikipedia webpage retrieval task, \ourmodel{} demonstrates significant improvements over the traditional text-based retrieval method BM25. 
Specifically, \ourmodel{} achieves 46.2\% and 77.6\% in top-1 and top-20 retrieval accuracy, which are 17 points and 10 points higher than BM25, respectively. 
This indicates that \ourmodel{} can effectively encode text-intensive documents in the format of screenshots for retrieval. 
When compared with neural text retrieval methods, \ourmodel{} outperforms smaller model DPR and performs on par with E5. Phi-3, which uses the same language model as \ourmodel{} (with 4 billion parameters), achieves approximately 4 points higher top-1 retrieval accuracy than \ourmodel{}. 
This suggests that existing vision language models still cannot fully capture the text content in a screenshot.

In the slide retrieval task, where the documents include a mix of text and visual content, we observe \ourmodel{} significantly outperforms (i.e., over 15 points in both nDCG@10 and Recall@10) all the text retrieval baselines that rely on OCR content extraction. 
This highlights the risk of information loss in the content extraction step, where OCR is only able to extract text content, thereby losing the visual elements of the documents. 
Notably, DPR, a neural retrieval method, fails to outperform BM25 in this task. 
This may be due to the varied layouts of slides, which pose additional challenges for text content extraction and result in noisy text input for text neural retrieval fine-tuning. 
By contrast, \ourmodel{} bypasses the stage of text content extraction and directly encodes document screenshots, which preserves more information for retrieval.

Finally, \ourmodel{} outperforms CLIP even though they use the same backbone of the vision transformer to digest the document screenshots.
For NQ, \ourmodel{} surpasses CLIP by 11.1 points in top-1 accuracy, and for SlideVQA, \ourmodel{} achieves 12.6 points higher in nDCG@10. 
We contribute the effectiveness gain to the large vision-language model encoder, which as we will show in Section~\ref{sec:impacts_of_patch}, has the capacity to handle more fine-grained information in a screenshot and possibly enhanced semantic understanding.

% \red{In order to know... , we studied hybrid retrieval results of text input method and screenshot input method as shown in Appendix~\ref{app:hybrid}.
% The results show that combining CLIP and text-based models leads to performance gains.
% However, DSE demonstrates stronger effectiveness compared to these hybrid results in SlideVQA, underscoring the advantages of our approach in encoding documents with integrated information and fine-grained visual understanding. We put it in Appendix because...}
To further explore the integration of text and visual information, we examined the hybrid retrieval results combining text-based and screenshot-based methods, as shown in Appendix~\ref{app:hybrid}. The results indicate that combining CLIP with text-based models yields notable performance improvements in the SlideVQA task.
However, DSE still outperforms such case in mixed modality scenario, demonstrating its capability to encode both fine-grained visual details and textual content directly in a single pipeline. 
As the hybrid approach is not a single, unified pipeline that directly encodes the document input, we leave the hybrid results in Appendix.

\begin{table}[t]
\centering\resizebox{1\columnwidth}{!}{
\begin{tabular}{l|cc|cc}
\toprule
\textbf{Zero-Shot} & \multicolumn{2}{c|}{\textbf{TriviaQA} } & \multicolumn{2}{c}{\textbf{SlideVQA-open} } \\
\textbf{Retriever} & Top 1 & Top 10 & nDCG@10 & Recall@10\\
\midrule
BM25 & 47.4 & 71.0 & 55.8 & 63.7 \\
% \hline
DPR & 37.3 & 65.5 & 29.5 & 39.7 \\
E5 & 46.9 & 73.1 & 42.6 & 54.4\\
Phi-3 & 57.1 & 78.1 & 49.7 & 62.1 \\
\midrule
CLIP & 37.3 & 65.6 & 48.4 & 61.6 \\
\ourmodel{} & 50.3 & 75.2 & 64.0 & 76.1 \\
\bottomrule
\end{tabular}}
\caption{Zero-shot retrieval effectiveness comparison. 
Models are trained on Wiki-SS with NQ questions and evaluated on TriviaQA questions and slide retrieval task.}
\label{tab:zeroshot}
\end{table}

\subsection{Zero-Shot Retrieval Effectiveness}
In this section, we further evaluate the generalization capability of \ourmodel{}.
Specifically, we apply the models fine-tuned on NQ questions to retrieve answers for TriviaQA questions~\cite{joshi-etal-2017-triviaqa} over the Wiki-SS (or the corresponding Wiki text) corpus, assessing their ability to generalize across different query distributions. 
Additionally, we evaluate the NQ fine-tuned models on the SlideVQA dataset to examine cross-task generalization.

As shown in Table~\ref{tab:zeroshot}, on TriviaQA, the text retriever based on LLM (i.e., Phi-3) achieves the best zero-shot effectiveness with a top-1 retrieval accuracy of 57.1\%. 
Both DPR and CLIP show lower zero-shot effectiveness, being outperformed by BM25 by approximately 10 points. 
In contrast, \ourmodel{} achieves a top-1 retrieval accuracy of 50.3\%, which is 3 points higher than BM25. 
This indicates that \ourmodel{} has relatively good zero-shot effectiveness across different query distributions but with room for improvement.

On the slide retrieval task, we observe that \ourmodel{} shows the best effectiveness among all. 
Specifically, \ourmodel{} outperforms BM25 by 8 points in terms of nDCG@10, while all the other text-based methods underperform BM25. 
This result shows that even though \ourmodel{} is only fine-tuned on the Wikipedia webpage retrieval task, where text is the main content, it is still able to encode document information beyond text. 
This demonstrates the potential of \ourmodel{} in handling diverse document types and tasks without needing task-specific training.

\begin{figure}[t]
    \centering
    \includegraphics[width=0.45\textwidth]{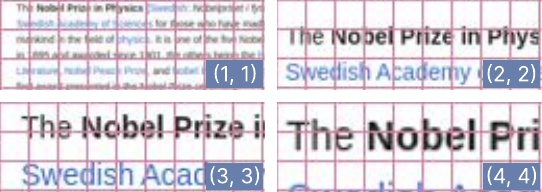}
    \caption{A snapshot of a Wikipedia webpage divided by different numbers of patches (red small squares). 
    As the number of patches increases, each patch can capture more fine-grained text information in the screenshot. 
    $(C_x, C_y)$ means the image are divided into $C_x \times C_y$ sub-images; then converted into $(C_x\times24)\times(C_y\times24)$ patches. 
    See more detail in Section~\ref{sec:method} and Figure~\ref{fig:enter-label}.
    }
    \label{fig:pixel}
\end{figure}

\subsection{Impacts of Patch Sequence Length}

As discussed in Section~\ref{sec:method}, each screenshot is cropped into $C_x \times C_y$ sub-images and encoded as a sequence of patches. 
Thus, increasing the number of crops yields a more lengthy patch input sequence, which incurs more computation cost for document encoding. 
On the other hand, increasing the number of crops results in patches with more fine-grained visual information, as illustrated in Figure~\ref{fig:pixel}. 
In the setting of $(C_x , C_y) = (1,1)$, each patch contains multiple words, while in the setting of $(C_x , C_y) = (4,4)$, a single letter is covered by two patches.
This leads to a trade-off between the efficiency and quality of document encoding. 
We study this trade-off by training \ourmodel{} with different numbers of crops and evaluate the corresponding retrieval effectiveness and document encoding speed (Doc/sec) on the Wiki-SS task for NQ questions.

\begin{figure}[t]
    \centering
    \includegraphics[width=0.45\textwidth]{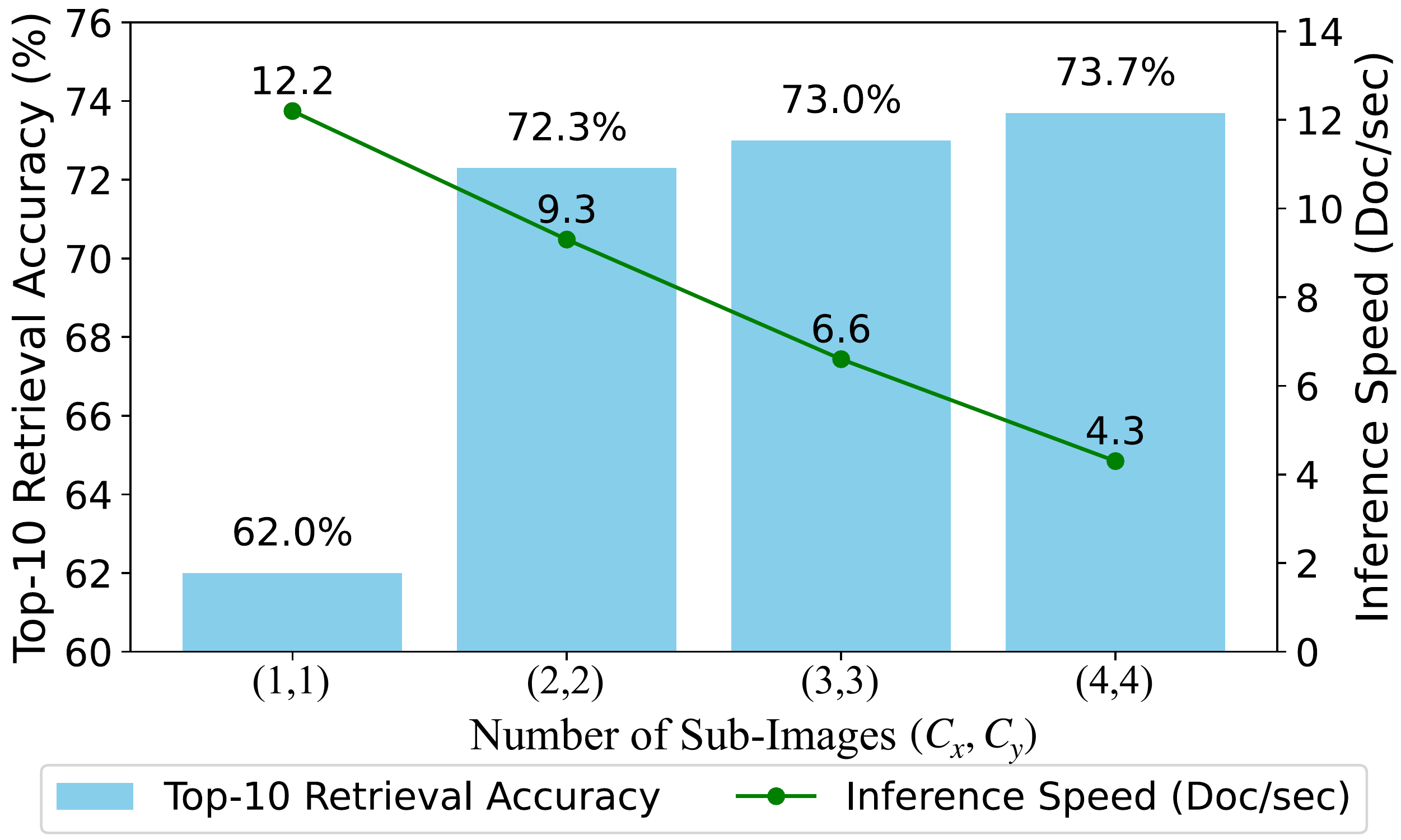}
    \caption{Trade-off between effectiveness and efficiency of \ourmodel{} with varying numbers of crops for input images. The inference speed is measured on a single H100 GPU with BF16 precision and FlashAttention enabled.}
    \label{fig:trade-off}
\end{figure}

We plot the efficiency and effectiveness in Figure~\ref{fig:trade-off}. 
When cropping the image into $4\times4$ sub-images for more fine-grained patch encoding, the top-10 retrieval accuracy increases from 62.0\% to 73.7\%, indicating that finer granularity helps the model better understand and encode the document screenshot.
However, this comes at the cost of computational efficiency. 
As the number of sub-images increases, the sequence length of the model's input grows, resulting in longer encoding times. 
The document encoding speed decreased from 12.2 documents per second with $1\times1$ sub-images to 4.3 documents per second with 
$4\times4$ sub-images as input. 
Finally, the experiment suggests that using $(C_x, C_y) = (2, 2)$ or $(3, 3)$ offers a good trade-off between retrieval effectiveness and computational efficiency of document encoding.

\label{sec:impacts_of_patch}

% \subsection{\red{Hybrid Search}}

\subsection{Analysis}

\subsubsection{Case Study}
\begin{figure}[t]
    \centering
    \includegraphics[width=0.45\textwidth]{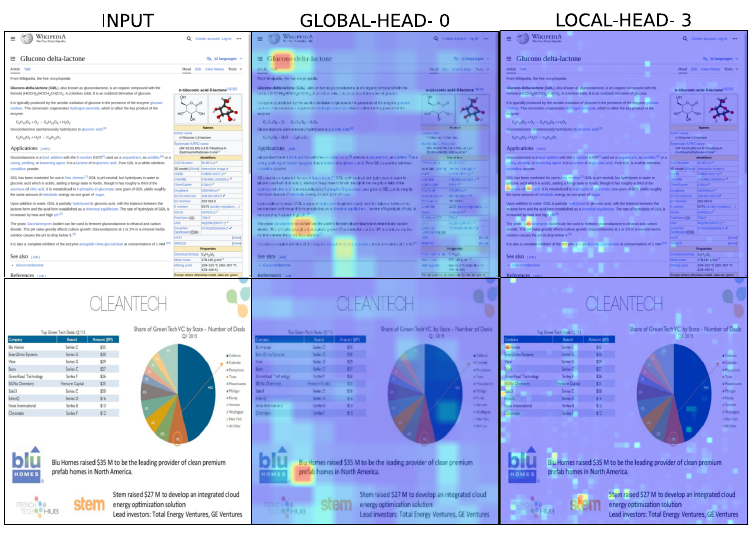}
    \caption{Case study on two examples in Wikipedia and SlideQA. We visualize the multi-head attention from the fine-tuned embedding to the image patches at the last layer. GLOBAL-HEAD is the attention head to the coarse image features (336$\times$336), while the LOCAL-HEAD is the attention head to more fine-grained image features after cropping (16$\times$336$\times$336).
    % We verify that the textual information is indeed extracted from the screenshots.
    % \red{shrink the case study to single column}
    }
    \label{fig:visualization}
    \vspace{-0.1cm}
\end{figure}

We conducted a case study to illustrate whether the fine-tuned embeddings effectively utilize the core semantic information in the screenshots. Figure~\ref{fig:visualization} presents the attention visualization of two examples from Wiki-SS and SlideVQA.
We used the Phi-3-vision model fine-tuned on NQ as the backbone and extracted the multi-head attention of the last token embedding to the image patches at the final layer.
The image patches contain both global and local features: Global features are tokenized from the resized full image input ($336\times336$), while local features are derived from crops when the image is resized to $1344\times1344$ and then cropped into $4\times4$ sub-images before encoding.
For both examples, the global attention heads appear to focus on general information, such as images, logos, titles, and sections.
In contrast, the local attention heads concentrate on finer details in the screenshots, such as individual letters and keywords, which are crucial for retrieval.
This qualitative evidence suggests that \ourmodel{} can effectively capture information from various modalities within the screenshots.

\begin{figure*}[t]
    \centering
    \includegraphics[width=\textwidth]{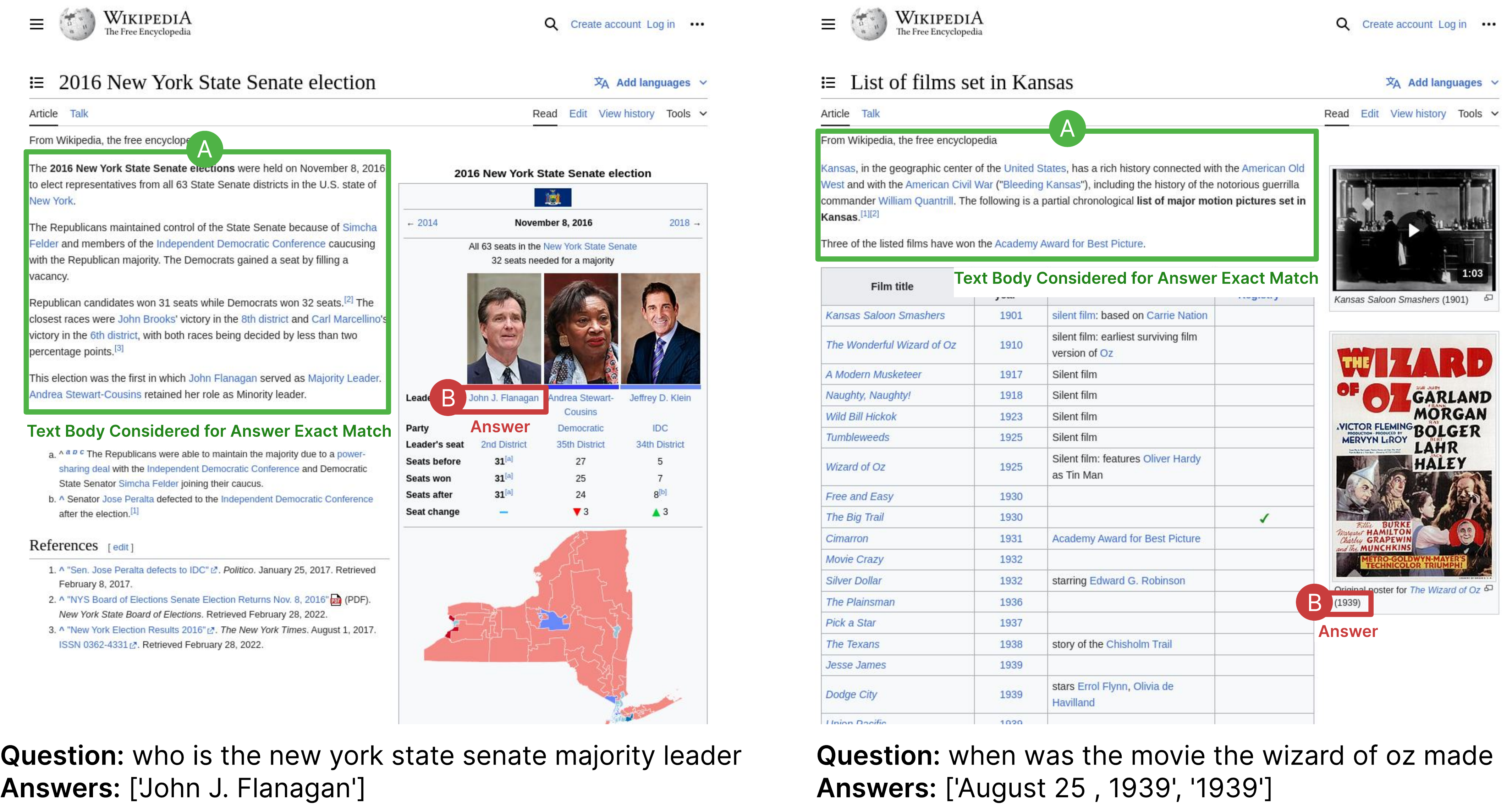}
    \caption{Examples of Top-1 retrieval results from \ourmodel{} for NQ test set questions that are being considered ``irrelevant'' because an exact match for the answer was not found in the corresponding extracted text body. However, the exact answer can be found in the tables covered by the screenshots.}
    \label{fig:error_analysis}
    \vspace{-0.1cm}
\end{figure*}

\subsubsection{Importance of Visual Integration}
In the mixed-modality retrieval task, both content extraction errors and the lack of visual context are inherent challenges in OCR-based methods, which our proposed \ourmodel{} method can overcome.
We conducted an error analysis of failure cases from the Phi-3 text retriever in the SlideVQA task, where \ourmodel{} retrieved relevant documents within the top 10 results, but Phi-3 did not.
We categorized the errors into two groups: (1) documents that could be answered using text alone, suggesting OCR errors, and (2) documents requiring additional visual context, indicating that missing the visual elements led to retrieval failures.
In our manual review of 50 cases, 22 could be resolved with correct text extraction, while 28 required visual context.
This analysis supports our claim that traditional OCR-based methods suffer from content extraction errors and loss of visual integration, while \ourmodel{} successfully addresses these issues by integrating all modalities.

\subsubsection{False-Negatives in Wiki-SS}
As mentioned in Section~\ref{sec:exp:web}, to evaluate \ourmodel{}, we examine whether there are any exact matches of the answer string in the retrieved documents. However, such evaluation only calculates the exact answer matches within the main text body. This could result in an underestimation of \ourmodel{}'s effectiveness if the answer appears in the content beyond the main text body, such as images, captions, or tables.
To investigate this potential underestimation, we randomly select 50 questions from the test set of NQ where \ourmodel{}'s top-1 retrieved documents are judged irrelevant while the purely text-based Phi-3 counter-part deems them positive.
We manually examine the corresponding screenshots retrieved by \ourmodel{} and discover that 7 out of 50 samples are actually false negatives.
In other words, the exact answer in these cases could be found in the image captions or tables within the screenshots as illustrated in Figure~\ref{fig:error_analysis}. This indicates \ourmodel{}'s capability to capture information in other areas besides the main texts that contain important clues for the document representation.

\section{Conclusion}
In this paper, we introduce \ourmodel{}, a novel information retrieval paradigm that leverages screenshots to simplify the document retrieval process. By circumventing traditional preprocessing steps and directly encoding documents with a vision-language model, \ourmodel{} offers a unified approach to handling varied document modalities.
We empirically show that \ourmodel{} outperforms traditional retriever and OCR-based methods on varied document retrieval tasks, such as webpage and slide retrieval.
This highlights the potential of \ourmodel{} to improve document retrieval in a range of real-world applications.

% Future developments could refine encoding techniques and adapt to different document types, setting new standards for multi-modal information retrieval.
By integrating \ourmodel{} with a large vision-language model (VLM) generator, it leads to a promising visual-based retrieval augmented generation (V-RAG) paradigm.
In this paradigm, \ourmodel{} retrieves document screenshots, which the VLM generator can process directly for generation, without needing separate text or image extraction.
This creates an end-to-end document intelligence system that eliminates the need for content extraction.
We hope our work opens future research into improving V-RAG with better retrieval methods, enabling more efficient and seamless multi-modal information retrieval and generation.

\section{Limitations}
This work has several limitations that warrant further exploration. Firstly, while we evaluated \ourmodel{} on Wikipedia webpage retrieval and slide retrieval datasets, there remains a gap in its effectiveness for more general-purpose document retrieval tasks, such as those involving PDFs or web pages with highly varied structures and content. Future work can consider multi-task training across diverse document types and content. Additionally, combining our method with extracted text and image contents could make \ourmodel{} more versatile for general retrieval tasks.
Secondly, our current approach relies solely on supervised fine-tuning. However, research in text retrieval has shown that contrastive pretraining can significantly improve retriever effectiveness. Investigating whether such pretraining methods can enhance \ourmodel{}'s performance is a promising direction for future research.
Thirdly, the reliance on visual data introduces challenges in environments where such data is of low quality. Blurry or low-resolution screenshots may degrade the effectiveness of \ourmodel{}. Conversely, processing very high-resolution images can reduce computational efficiency. We leave further explore the balance of image quality and computational efficiency as future work.
% \red{Lastly, this work focus on English retrieval task, how such method applies to non-English especially low resource languages requires further exploration.}

\section*{Ethics Statement}
This work complies with the ACL Ethics Policy.
We declare that there are no ethical issues in this paper, to the best of our knowledge.

\section*{Acknowledgments}
We sincerely thank Xilun Chen, Xinyu Shi, Xinyu Zhang, Dawei Zhu, and the anonymous reviewers for their invaluable feedback and insightful suggestions.
We also extend our appreciation to Jheng-Hong Yang, Dongfu Jiang, and Yobo Wang for their helpful discussions on technical questions.
This research was supported in part by the Natural Sciences and Engineering Research Council (NSERC) of Canada and Microsoft via the Accelerating Foundation Models Research program.

% \clearpage

\bibliography{latex/main}

\appendix
\section{Appendix}
\label{sec:appendix}

\subsection{Wiki-SS Context Length}
\label{app:wiki-ss}
For Wiki-SS, the screenshot size is consistently $980\times980$. 
However, it is challenging to ensure that the text version exactly matches the text body in the screenshot.
We manually check a sample of 50 Wikipedia pages where the main content body was longer than what was covered in the first-page screenshot.
The maximum number of words covered in these cases was 492, with an average of 453.
Therefore, we set the truncation length to 500 words to ensure that the text version does not contain less content than the screenshot so that \ourmodel{} does not benefit from having more main text content when demonstrating its effectiveness.

Additionally, we examined the sensitivity of the text-based retriever regarding input length in the Wikipedia retrieval task. The table below shows the effectiveness of the text retriever E5 on the Wiki-NQ task with varying input lengths:

\begin{table}[!h]
\centering
\begin{tabular}{lcc}
\toprule
\textbf{Max Len.} & \textbf{Top-1 Acc.} & \textbf{Top-5 Acc.} \\
\midrule
500 & 47.64 & 68.58 \\
450 & 47.12 & 68.36 \\
400 & 47.47 & 68.00 \\
300 & 47.34 & 67.14 \\
200 & 45.87 & 66.70 \\
100 & 43.60 & 64.73 \\
\bottomrule
\end{tabular}
\caption{Effectiveness of the E5 text retriever on the Wiki-NQ task with varying input lengths.}
\label{tab:length}
\end{table}

\noindent As shown in Table~\ref{tab:length}. The top-k retrieval accuracy converges when the input length exceeds 400 words, indicating that setting the truncation point at 500 provides a reliable evaluation without unfair advantages or disadvantages.

\subsection{Text--Screenshot Hybrid Search}
\label{app:hybrid}

We evaluate whether combining text-based input methods with visual-based input can improve retrieval performance.
Specifically, we examine the effectiveness of a hybrid search pipeline that incorporates both text-based retrieval and screenshot-based retrieval. The hybrid approach is achieved by interpolating the similarity scores of the ranking results from the two retrievers~\cite{hybrid}.
% , following the strategy used in \cite{hybrid}.

As shown in Table~\ref{tab:hybrid}, the hybrid CLIP and Phi-3 search significantly improves performance on the SlideVQA task, outperforming each method individually, with notable gains in nDCG@10 and Recall@10. This suggests that integrating text and visual information provides a more comprehensive understanding, enhancing retrieval in mixed-modality scenarios, especially when CLIP alone struggles with text content.
However, for the text-intensive task (Wiki-SS), we find that hybrid models do not offer significant improvements. However, for the text-intensive Wiki-SS task, hybrid models show minimal improvement indicating limited benefits of hybrid input in such context.

Interestingly, while the hybrid approach boosts performance for mixed-modality tasks, \ourmodel{} alone still outperforms the hybrid CLIP + Phi-3 model, highlighting its ability to directly integrate both text and visual information effectively.

\subsection{Dataset Licenses}
\begin{itemize}
    \item \textbf{NQ}: Apache License 2.0
    \item \textbf{TriviaQA}: Apache License 2.0
    \item \textbf{SlideVQA}: SOFTWARE LICENSE AGREEMENT FOR EVALUATION
    \item \textbf{Wikipedia}: Creative Commons Attribution Share Alike, GNU Free Documentation License family.
    \item \textbf{Wiki-SS}: Creative Commons Attribution Share Alike, GNU Free Documentation License family.
\end{itemize}

\begin{table*}[t]
\centering
\begin{tabular}{l|c|cccc|cccc}
\toprule
\textbf{Retriever} & \textbf{Document}& \multicolumn{4}{c|}{\textbf{NQ}} & \multicolumn{2}{c}{\textbf{SlideVQA-open}} \\
\textbf{} & & Top 1 & Top 5 & Top 10 & Top 20 & nDCG@10 & Recall@10\\
\midrule
Phi-3 & Text & 50.6 & 70.9 & 75.8 & 79.5 & 59.0 & 69.5 \\
\midrule
CLIP & \multirow{2}{*}{Screenshot} & 35.1 & 57.7 & 64.8 & 71.2 & 61.7 & 74.7 \\
\ourmodel{} &  & 46.2 & 68.5 & 73.7 & 77.6 & 75.3 & 84.6 \\
\midrule
CLIP + Phi-3 & \multirow{2}{*}{Hybrid} & 50.2 & 69.5 & 75.1 & 78.9 & 67.9 & 80.5 \\
\ourmodel{} + Phi-3 & & 51.7 & 71.2 & 75.8 & 79.1 & 71.5 & 83.7 \\
\bottomrule
\end{tabular}
\caption{Effectiveness of retrieval effectiveness by fusing ranking the ranking results of text-input method and screenshot-input method.}
\label{tab:hybrid}
\vspace{-0.2cm}
\end{table*}

\subsection{Hyper-Parameters for Training}
\label{sec:hyper}
\begin{figure*}[t]
    \centering
    \includegraphics[width=\textwidth]{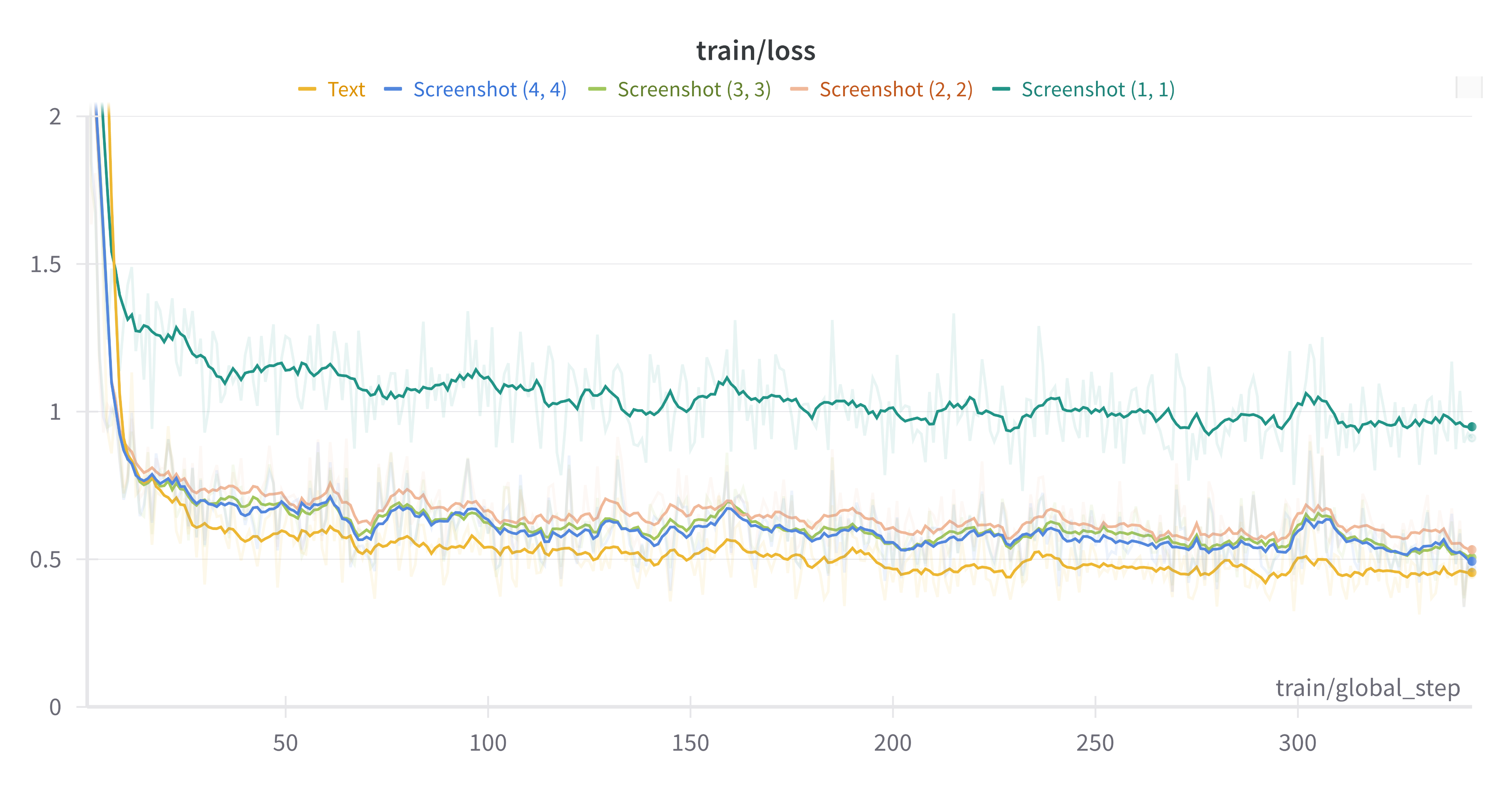}
    \caption{Loss curve of training \ourmodel{} variants.}
    \label{fig:loss_curve}
    % \vspace{-0.2cm}
\end{figure*}

Please see Table~\ref{tab:hyper} for detailed hyper-parameters setting. Additionally, loss curve for training \ourmodel{} models can be found at Figure~\ref{fig:loss_curve}.

\subsection{AI Assistant Usage}
GPT4o is used during the writing to correct grammar errors and format tables.

\begin{table*}[t]
\centering
\small
\resizebox{0.95\textwidth}{!}{
\begin{tabular}{|l|l|l|l|l|l|}
\toprule
\textbf{Method} & \textbf{DPR} & \textbf{E5} & \textbf{Phi3} & \textbf{CLIP} & \textbf{\ourmodel{}} \\ 
\midrule
Model Init & google-bert/bert- & intfloat/e5-base- & microsoft/Phi-3- & openai/clip-vit-large- & microsoft/Phi-3- \\
     & base-uncased      & unsupervised      & vision-128k-instruct & patch14-336 & vision-128k-instruct \\
License & Apache 2.0 & MIT License & MIT License & MIT License & MIT License \\ 
\# of Parameters & 110 M & 110 M & 4B & 430 M & 4B \\ 
Backbone Modality & text & text & text or vision & text XOR vision & text OR vision \\ 
Learning Rate & 1e-5 & 1e-5 & 1e-4 & 1e-5 & 1e-4 \\ 
GPU & 2xA100 80G & 2xA100 80G & 2xA100 80G & 2xA100 80G & 2xA100 80G \\ 
Per Device Batch Size & 64 & 64 & 8 & 16 & 8 \\ 
Hard Neg Per Query & 1 & 1 & 1 & 1 & 1 \\ 
Gradient Accumulation & 1 & 1 & 8 (4) & 4 & 8 (4) \\ 
Total Batch Size & 128 & 128 & 128 (64) & 128 & 128 (64) \\ 
Pooling & cls & mean & eos & mean & eos \\ 
Temperature & 1 & 0.02 & 0.02 & 0.02 & 0.02 \\ 
Normalize & False & True & True & True & True \\ 
Epochs & 40 & 40 & 1 (2) & 10 & 1 (2) \\ 
LoRA & False & False & True & False & True \\ 
LoRA r & N/A & N/A & 8 & N/A & 8 \\ 
LoRA Alpha & N/A & N/A & 64 & N/A & 64 \\ 
LoRA Dropout & N/A & N/A & 0.1 & N/A & 0.1 \\ 
LoRA Target & N/A & N/A & \text{*\_proj} & N/A & \text{*\_proj} \\ 
\bottomrule
\end{tabular}}
\caption{Detailed hyper-parameter settings for baselines and our method. By default, the parameters are for the Wiki-SS NQ training. If the setup is different for SlideVQA training, it is noted in parentheses.}
\label{tab:hyper}
\end{table*}

% \label{sec:appendix}

% This is an appendix.

\end{document}